\documentclass[11pt]{article}

\usepackage{lipsum} 
\usepackage{amsmath}
\usepackage{amssymb}
\usepackage{amsthm}
\usepackage{epsfig}
\usepackage{algorithm,algorithmic}
\usepackage{mathtools}
\usepackage{graphicx}

\usepackage[sc]{mathpazo} 
\usepackage[T1]{fontenc} 
\usepackage{microtype} 

\usepackage[hmarginratio=1:1,top=28mm,columnsep=20pt]{geometry} 
\usepackage{multicol} 
\usepackage[hang, small,labelfont=bf,up,textfont=it,up]{caption} 
\usepackage{booktabs} 
\usepackage{float} 
\usepackage{hyperref} 

\usepackage{lettrine} 
\usepackage{paralist} 

\usepackage{abstract} 

\usepackage{titlesec} 
\renewcommand\thesection{\Roman{section}} 
\renewcommand\thesubsection{{\Roman{subsection}}} 
\renewcommand\thesubsubsection{{\Roman{subsubsection}}} 
\titleformat{\section}[block]{\LARGE\scshape\centering\bfseries\textbf}{\thesection.}{1em}{} 
\titleformat{\subsection}[block]{\Large\centering\bfseries\textbf}{\thesubsection.}{1em}{} 
\titleformat{\subsubsection}[block]{\large\centering\bfseries\textbf}{\thesubsubsection.}{1em}{} 

\newtheorem{theorem}{Theorem}

\newtheorem{definition}[theorem]{Definition}

\usepackage{fancyhdr} 
\title{\vspace{-15mm}\fontsize{22pt}{10pt}\selectfont\textbf{Twitter Hash Tag Recommendation}} 

\author{
\large
\textsc{Roman Dovgopol and Matt Nohelty}\\ 
\normalsize University of Minnesota \\ 
\vspace{-5mm}
}
\date{}

\begin{document}

\maketitle 

\thispagestyle{fancy} 


\begin{abstract}

\noindent The rise in popularity of microblogging services like Twitter has led to increased use of content annotation strategies like the hashtag. Hashtags provide users with a tagging mechanism to help organize, group, and create visibility for their posts. This is a simple idea but can be challenging for the user in practice which leads to infrequent usage. In this paper, we will investigate various methods of recommending hashtags as new posts are created to encourage more widespread adoption and usage. Hashtag recommendation comes with numerous challenges including processing huge volumes of streaming data and content which is small and noisy. We will investigate preprocessing methods to reduce noise in the data and determine an effective method of hashtag recommendation based on the popular classification algorithms.

\end{abstract}


\section{Introduction}

\noindent 
Microblogging services have become increasingly popular in recent years allowing users to share small snippets of text with the world. Users share a variety of information from personal opinions and thoughts to breaking news and product marketing messages. Of the many microblogging services available, Twitter has become the dominant service with estimates of 250 million registered users who post up to 500 million Tweets per day [12].
\\

\noindent 
As Twitter grew in popularity, it became apparent that an organizational strategy was needed to organize the massive amounts of data. In 2009, Twitter introduced a user driven annotation called a hashtag [3]. Today, hashtags are used for many purposes such as finding posts of interest, organizing a group message, and determining the most popular topics of discussion. Users can apply hashtags to their posts by placing the number sign (\#) before a word they want to use as a hashtag. Users are free to create almost any hashtag they can imagine as long as it is a single alphanumeric string of text. The practical ramifications of this mean that hashtags containing more than one word are simply pushed together by removing spaces so these hashtags are not a single correctly spelled word. For example, a user who enjoys big data processing might Tweet something like "I heart \#bigdataprocessing".
\\

\noindent 
Hashtags are simple to apply and well known to mainstream Twitter users. However, hashtags are only used in approximately 10\% of Tweets [4]. This limited usage greatly reduces the usefulness of the hashtag as a global organizational strategy. If hashtags could be recommended to users before they post their Tweet, we suspect hashtag usage would increase dramatically. 
\\

\noindent 
In this paper, we develop a method for recommending relevant hashtags to users in real-time. Hashtag recommendation has two main challenges that set it apart from traditional document tag recommendations. First, the content of a Tweet is very short and often includes abbreviations, misspellings, and incorrect grammar. The limited number of words in a Tweet makes traditional document classification techniques such as TF-IDF (Term Frequency and Inverse Document Frequency) ineffective because of the TF=1 challenge which means that words are rarely repeated in a Tweet [10]. This makes it difficult to determine which words are most important to the meaning of the Tweet. The second challenge is the sheer volume and speed at which data is produced. Tweets are produced constantly so the incomming Tweets need to be treated as an infinite length stream. Our approach to recommendation attempts to handle these challenges with a variety of preprocessing techniques and a recommendation model based on the popular classification algorithms K-Nearest Neighbor and Naive Bayes.


\section{Constraints}

\noindent 
Three main constraints were taken into account when developing our recommendation system. First, our recommendations are based solely upon the content of the Tweet. We scoped the problem this way because the content of a Tweet must be the core factor of any hashtag recommendation system so it is essential to develop a solid approach using this factor alone before extending the recommender to consider other factors like the author's social network or previous posting history. 
\\

\noindent 
The second constraint is speed. For this appraoch to be useful in a real world application, we need to generate many recommendations very rapidly. Today's internet users are unwilling to wait for a web page for even a few seconds [14] so the usefulness would be greatly reduced if recommendations could not be produced in milliseconds.  Also, as previously mentioned, Twitter has seen as many as 500 million posts per day, so a recommendation system needs to be able to scale to handle large volumes of input.
\\

\noindent 
The third and final constraint is computational resource demands. The recommendation system needs to process and infinite length stream, run indefinitely, and scale gracefully as the stream progresses so demands on computional resources like CPU and memory need to remain relatively constant.


\section{Prior Work}

\noindent 
There has been significant research on keyword extraction, document classification, and tag recommendation but far less as it relates to very short documents like microblogging posts. In this section, we briefly review the main directions for the studies related to this paper.
\\

\noindent 
In some ways, hashtag recommendation is similar to a typical text classification problem and various algorithms from regression and classification analysis might be applied here [7][18]. Other studies focus on classification of short-text documents but commonly use only a limited number of features. Therefore they almost always degrade if we consider each hashtag as a separate class. The main reason for this is the obvious obstacles of maintaining and processing an exceedingly big number of classes [2][8].
\\

\noindent 
Other related methods for hashtag recommendation are based on measuring the similarity of tweets. Zangerle[5] and Godin[6] have several recommendation approaches based on TF-IDF methods and feature vectors for each tweet. Also, several papers have been published on recommending hashtags by using external meta data sources such as WordNet and Wikipedia to build context which could be applied to the text of post [11][15]. A study by Kharibi[4] considers a tweet as a set of words and measures a relevance between the words by aggregating all relevance scores which are computed from a co-occurrence graph. Feng's research [16] develops and expands this approach both with variety of optimizations and improvements and by processing a user's personal information. In this paper, we apply concepts from these papers to develop our own hashtag recommendation approach.


\section{Our Approach}

\noindent
Our hashtag recommendation process can be broken down into a three main topics: preprocessing, classification, and computational resource management. 


\subsection{Preprocessing}

\noindent
Preprocessing is extemely important in many big data applications because it provides a mechanism to filter, aggregate, and cleanse the dataset making the actual processing more accurate and efficient. Hashtag recommendation is one such application that benefits greatly from this approach. Tweet content is very noisy when compared to many other document datasets because character constraints force abbreviations and posts are written by millions of different people so there is no consistency between authors.  Because of this, preprocessing Tweets is an essential first step to minimize the impact of this noise to prepare the data for accurate classification.
\\

\noindent
The first preprocessing step was to normalize the character set by converting all alpha characters to lower case and stripping out punctuation. Our classifiers work by finding terms that appear in multiple Tweets so this normalization step allows us to group words with the same meaning. For example, "Don't", "don't", and "dont" are all the same word but are written differently. This normalization will convert each of these words to "dont" so each occurrence will be grouped together making the distribution of terms more favorable for processing and more reflective of the intent of the authors.
\\

\noindent
For our classifiers to work most effectively, they need to base the classification on words with the most significance to the meaning of the Tweet. To maximize the likelihood of this, we removed words that have a high probability contributing little to the meaning of the Tweet. We removed 175 common English stop words like "a", "the", and "and" as well as common words found on Twitter like "rt" (i.e. retweet). We also removed any words that contained less than 3 characters and words made up entirely of numbers. We also exclude links from consideration in our classifiers. Because of the 140 character limit, links in Twitter are generally created using url shortening tools like bit.ly (https://bitly.com) which generates unique links for each user even if they point to the same destination page. Because of this, very little information can be gleaned from the url.
\\

\noindent
The final preprocessing step we investigated was stemming using the Porter Stemming algorithm [9] which removes prefixes and suffixes from each word in a Tweet. The intution was that this would improve the accuracy of the classifier because terms with a common stem usually have similar meanings [9]. In practice however, we found this to have neglible impact on the accuracy of our recommendations as well as a non-trivial decrease in recommendation speed. Because of this, we excluded this preprocessing step in our final algorithm. We did not investigate why stemming did not improve accurracy but our suspicion is that the many abbreviations, mispellings, and slang terms commonly used on Twitter interfere with the stemming algorithm. Further research is necessary here to determine if modifications to the stemming algorithm could improve results.


\subsection{Classifications}

\noindent 
Our recommendation engine is based primarily on classification. We evaluated a variety of clustering and classification algorithms and selected K-Nearest Neighbor and Naive Bayes to base our recommendations on. These algorithms are straight forward to implement but more importantly, they are lazy evaluating algorithms. This means the underlying data of the classification model can be modified without needing to rebuild the entire model. By contrast, decision trees and rule based classifiers require the decision tree or rule sets to be regenerated if the model data changes.   


\subsubsection{Naive Bayes}

\noindent 
Naive Bayes classification uses Bayes' Theorem to determine the conditional probability of a class given an item's features. In this application, the class is the hashtag and the features are the words in a Tweet. Bayes Theorem makes it possible to calculate the probability of a hashtag given a set of words in a Tweet. If this is done for all known hashtags, the hashtags with the highest probability can be used as the recommendations.

\begin{theorem}[Bayes' Theorem] Multiple Features
 \begin{eqnarray*} P(h|w_{1},w_{2},...,w_{n}) = \frac{P(w_{1},w_{2},...,w_{n}|h) *  P(h)}{P(w_{1},w_{2},...,w_{n})} = P(h|w_{1}) * P(h|w_{2}) * ... * P(h|w_{n}) \end{eqnarray*}
\end{theorem}

\noindent 
However, there are a few problems with this approach. In many cases, $P(h|w_{i})$ will be zero when we have not yet seen a Tweet containing word $w_{i}$. If this occurs, the entire $P(h|w_{1},w_{2},...,w_{n})$ goes to zero even if other words have high conditional probabilities. We investigated methods such as Laplace Smoothing which did improve our results, but we found that summing the conditional probabilities instead of using the product produced even better results. The intuition behind this is that in many cases a single word is a strong predictor of the hashtag. If most of the words in a Tweet are unrelated to a hashtag but one word is very related, the unrelated words drag down the overall probability of that hashtag when computing the product. By summing the conditional probabilities, we preserve the high weighting contributed by the high probability words. Summing also handles the issue where $P(h|w_{i})$ is zero because a word with a score of zero has no negative impact on $P(h|w_{1},w_{2},...,w_{n})$.
\\

\noindent 
Below is our optimal Naive Bayes algorithm. Let $W$ be the list of words in the Tweet we are classifying. Let $T$ be the list of Tweets in our classification model. Let $n$ be the number of hashtag classifications to return.
\\

\begin{algorithm}
\begin{algorithmic}
\STATE $T_{s} = SIMILAR\_TWEETS(W, T)$
\STATE $H = T_{s}.hashtags$ //All hashtags in list of Tweets
\STATE $classifications$ = [ ]
\FOR {$h$ in $H$}
  \STATE $score$ = 0
  \FOR {$w$ in $W$}
     \STATE $score$ += $(\frac{P(w|h) *  P(h)}{P(w)} * WORD\_WEIGHT(w))$
  \ENDFOR
  \STATE $classifications.add(h, score)$
\ENDFOR
\RETURN $getHighestScores(classifications,n)$ //Return n highest scored hashtags
\end{algorithmic}
\caption{[$NB\_CLASSIFY({W, T, n})$]}
\end{algorithm}


\subsubsection{K-Nearest Neighbor}

\noindent 
K-Nearest Neighbor classifiers find the K most similar data points to the item being classified and return the most common class found in those K items. The key to a successful K-Nearest Neighbor algorithm is the similarity measure to compare two items. There are many ways to do this from simple calculations like the Tanimoto Distance (number of words found in both documents) to weighting measures like TF-IDF (Term Frequency and Inverse Document Frequency). For the Tweet dataset, we found the most successful similarity measure to be the sum of the Term-Corpus Relevance (TCoR) [10] scores for all words found in both Tweets. TCoR is a weighting measure proposed by Timonen, et al. to measure how strong of a class predictor the word is across the entire dataset.  
\\

\begin{definition}
\begin{algorithmic}
\STATE $TCoR(w) = \frac{\frac{1}{fl(w)} + \frac{1}{c_{w}}}{2}$ 
\end{algorithmic}
\end{definition}

\noindent 
$fl(w)$ is the average number of words in Tweets containing the word and $c_{w}$ is the number of hashtags the word co-occurs with. The intuition for TCoR is that the fewer number of words in a Tweet, the more important an individual word is to the overall meaning of that Tweet and words occurring with a small number of distinct hashtags are more imformative when predicting a hashtag [10]. Timonen, et al. propose a second measure called Term-Category Relevance (TCaR) to combine with TCoR for the final weight. Interestingly, using TCaR reduced the accuracy of our recommendations by almost 50\% so we used TCoR alone to weight the occurrence of words. 
\\

\noindent 
Below is our optimal K Nearest Neighbor algorithm. Let $W$ be the list of words in the Tweet we are classifying. Let $T$ be the list of Tweets in our classification model. Let $K$ be the number of nearest neighbors to consider. Let $n$ be the number of hashtag classifications to return.
\\

\begin{algorithm}
\begin{algorithmic}
\STATE $T_{s} = SIMILAR\_TWEETS(W, T)$
\STATE $neighbors =$ [ ]
\FOR {$t$ in $T_{s}$}
     \STATE $score = TCoR(w) * WORD\_WEIGHT(w)$
     \STATE $neighbors.add([t, score])$
\ENDFOR
\STATE $nearest\_neighbors = getHighestScores(neighbors, K)$ //Get K tweets with highest score 
\STATE $classifications =$ [ ] //Map of hashtag and count
\FOR {$t$ in $nearest\_neighbors$}
     \FOR {$h$ in $t.hashtags$}
    	\STATE $classifications.add([h,1])$ //Add h with count of 1 (or increment count)
     \ENDFOR
\ENDFOR
\RETURN $getHighestScores(classifications, n)$ //Return n hashtags with highest count
\end{algorithmic}
\caption{[$KNN\_CLASSIFY({W, T, K, n})$]}
\end{algorithm}


\subsubsection{Shared Classifier Functions}

\noindent 
Both the Naive Bayes and K Nearest Neighbor classifiers share a few other enhancements to improve classification accuracy and speed. We could compute the classification by considering every Tweet in the dataset but that would be very slow as the dataset could contain millions of Tweets. Instead, we use $SIMILAR\_TWEETS(W, T)$ to quickly narrow down the full list of Tweets to just the most likely matches. 
\\

\noindent  
We do this by computing the Inverse Document Frequency (IDF) for all the words in $W$. IDF is a commonly used measure of how common a word is across a set of documents. It is an important measure because not all words contribute equally to the meaning of a document. A high IDF score indicates the word is rare across the document set which means it is likely important to the meaning of a document in which it is found. 
\\

\begin{definition}
\begin{algorithmic}
\STATE $IDF(w) = log(\frac{|T|}{|w \in T|})$ 
\end{algorithmic}
\end{definition}

\noindent 
We select the three words from $W$ with highest IDF scores as these are likely to be the three most important words in the Tweet we are classifying. We then find all Tweets in our dataset that contain at least one of those three words. This list of Tweets is just a fraction of the size of the entire Tweet dataset and likely contains Tweets most similar to the one we are classifying. Limiting our classification algorithms to investigate only this list of Tweets improves the speed of classifications and also helps to filter some of the noise created by high frequency words.
\\

\noindent
The second shared function is $WORD\_WEIGHT(w)$ which computes additional weighting for a given word. If a word starts with an '@' it is a mention, which means the Tweet is directed to a certain user. We apply a weight of 3 to all mentions because Tweets directed to the same user have a high likelihood of being similar and containing the same hashtag.  If the word is not a mention, we give it a weight of 1 plus 0.1 for every letter in the word. This weights longer words more heavily than shorter words. The intuition is that longer words generally have a more specific meaning and are more likely to be important to the meaning of the Tweet.

\subsubsection{Hybrid Classifier}
\noindent 
Our final classification model makes use of both the Naive Bayes and K Nearest Neighbors classifiers described above. We compute classifications using both algorithms and then combine the results with a weighting factor we determined experimentally. Each hashtag recommended by Naive Bayes recieves a weight of 0.4 and each hashtag recommended by K Nearest Neighbor receives a weight of 0.6. We sum up the weights and return the hashtags with the highest scores. Using both classifiers takes advantage of the strengths of each classifier which causes the most likely hashtags to move to the top of the list of recommendations.


\subsection{Computational Resource Management}

\noindent 
Finally, we need to consider speed and computational resources. Tweets are produced in high volume, 24 hours a day, so being able to produce recommendations quickly with constant computational resources is extremely important. Traditionally, classification algorithms are given a training set of data up front to train the classifier and all future classifications are computed using that initial model. This method doesn't work for a stream of Tweets because hashtag usage changes very rapidly on Twitter as topics trend and fade away. Because of this it's very important to keep the classifier's model as current as possible. 
\\

\noindent
We modified the classifiers to use the "sliding window" pattern which ensures the most recently seen Tweets are used in the model [4]. In our implementation, the classifier data model is a FIFO (first in first out) queue. First, the classifier is initialized with a predefined number of Tweets. Following initialization, every time a new Tweet arrives from the stream, the oldest Tweet is removed from the model and the new Tweet is added. Because we chose to use Naive Bayes and K Nearest Neighbor, which are lazy evaluating classifiers, adding and removing Tweets to and from the dataset is very fast because we do not need to regenerate any structures built on top of the model like a rule set or a decision tree. 
\\

\noindent 
However, because these are lazy evaluators, classification itself can take longer than algorithms that use complex prebuilt structures. Because of this, we do maintain some external structures that are easy to update when Tweets are added and removed from the model. Besides the ordered list of Tweets in the model itself, we keep two hash maps, one indexed by a word and the other indexed by a hashtag. Each of these hash maps store basic statistics about the word or hashtag which we use to compute the weights and probabilities needed in our classification algorithms. Because these are hash maps, we can find statistics about a given word or hashtag in constant time. This provides a nice compromise between lazy and eagar classifiers which allows for efficient updates to our data model as it changes but also supports fast computation of classifications.
\\

\noindent 
In order to further improve the speed of our recommendations, we parallelized our classifiers. We observed significant speed improvements when running our recommendations with multiple threads and did not observe any decrease in recommendation accuracy. This is an important finding because it shows our classification can scale to handle larger data volumes. We did not extend this parallelization to multiple machines so this would be an interesting area to investigate further.
\\

\noindent 
The final area we investigated to speed up classification was maintaining a seperate model for each language. This provides a great opportunity to parallelize the recommendation system even further. Each language could have it's own classifier and model which would function independently, and it would only need to process a fraction of the full volume of Tweets. For example, when an English Tweet is received it could be routed to the machine, or cluster of machines, that maintains the English model and performs English classifications. This system could be duplicated for each language. We proved out this architecture on a single machine and achieved classifications of comparable accuracy but found the language detection itself to be such a bottleneck that it drastically slowed down recommendations. It would be interesting to investigate this further to see if a faster language detection algorithm could be designed to take advantage of this further parallelization.


\section{Experimental Results}

\noindent
To evaluate our recommendation algorithm, we simulate how a system like this would function in a real world application. We begin by initializing the classifier by fully populating the sliding window dataset with Tweets with known hashtags. When initialization completes, we start producing recommendations for Tweets with known hashtags. For each Tweet, we recommend up to three hashtags. If one of those hashtags matches the actual hashtag of the Tweet, we consider this recommendation a success. Following each recommendation, we add that Tweet to the sliding window dataset and removed the oldest Tweet which maintains our sliding window.
\\

\noindent
We ran our evaluation with a sliding window size of 1 million Tweets and produced recommendations for an additional 1 million Tweets. For each Tweet, we chose to recommend up to three hashtags as this is a realistic number that could be shown to a user in a real application.  The dataset we used for testing contained a random sampling of 35,575,057 Tweets collected from 4/14/13 to 4/21/13.  Of that, 4,038,934 Tweets (11.3\%), contained hashtags.    
\\

\begin{figure}[ht!]
\centering
\includegraphics[width=90mm]{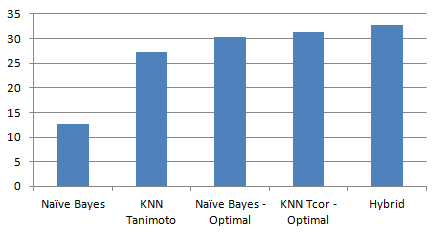}
\caption{Successful classification percentages for each classifier}
\label{success_chart}
\end{figure}

\noindent
The optimal Naive Bayes recommender successfully recommended hashtags for 30.22\% Tweets and the optimal K Nearest Neighbor recommender successfully recommended hashtags for 31.4\% of Tweets. When we combined these together in our Hybrid recommender we were able to achieve a successful recommendations for 32.71\% of the test Tweets. Figure 1 shows the success percentages for each classifier. We observed a precision of 0.2, recall of 0.27, and an F measure of 0.23.
\\

\noindent
This evaluation method is very conservative and underestimates the actual success rate of our recommendations for a few key reasons. First, each recommendation is considered successful only if the actual hashtag exactly matches one of the recommendations. In reality, most Tweets don't have one single correct hashtag but could have numerous hashtags that would make sense to a user and would be considered good recommendations by a human.  However, due to resource and time constaints, it was not possible to have a real person evaluate the millions of recommendations made by our classifiers during these evaluations. 
\\

\noindent
The second reason is that we exclude the known hashtag of the Tweet that we are making recommendations for.  Including the hashtag unfairly biases the recommendation in cases where the hashtag is truely a tag. For example, consider a Tweet like "Just heard a funny joke \#laughoutloud". In this case, the user would never have included "\#laughoutloud" if they did not know about that hashtag. Because of this, we removed hashtags from the Tweet as we did not want to give our recommender an unfair advantage.  However, this negatively impacts the recommendations in other cases where the hashtag is part of the content of the Tweet.  For example, "Just heard a funny \#joke".  In this case, the user would have likely typed in the word "joke" even if they did not know to use it as a hashtag. Because we are excluding hashtags, we would not use the word joke when producing recommendations for this Tweet so we are making this recommendation without an important word, making it far less likely for us to produce an accurate recommendation. Including the hashtag when making our recommendations nearly doubled our success rate.  
\\

\noindent
The Hybrid recommender processed the 1 million test Tweets in just under 29 minutes at an average rate of 576 Tweets / Second. This equates to approximately 50 million Tweets per day. These tests were run on a standard laptop. In a real production application, much more powerful hardware would be available that could fully take advanatage of the parallelization capabilities discussed previously so we would expect even better performance. 


\section{Conclusions}
\noindent
In this paper we discussed the challenges that come with Twitter hashtag recommendation and proposed solutions to overcome them. We combined these ideas to create a scalable recommender system based on Naive Bayes and K Nearest Neighbor classifiers. We used the sliding window pattern to maintain constant computational resource demands and used IDF to efficiently filter down the dataset to allow recommendations to be produced rapidly. To improve recommendation accuracy, we combined classifiers and used a variety of weighting measures to ensure these classifiers consider only the most significant words in each Tweet. Throughout the paper, we also proposed a few key areas to further extend our work to build upon the speed and accuracy we were able to achieve with this recommender system.
\\

\newpage
















\end{document}